\documentclass[prl,aps,superscriptaddress,twocolumn,amssymb,showpacs]{revtex4-1}

\usepackage{color}
\usepackage{graphicx}
\usepackage{longtable}
\usepackage{epsfig}
\usepackage{dcolumn}
\usepackage{bm}
\usepackage{amssymb}

\def\beq{\begin{equation}}
\def\eeq{\end{equation}}
\def\bea{\begin{eqnarray}}
\def\eea{\end{eqnarray}}

\begin{document}

\title{Order in Quantum Compass and Orbital $e_g$ Models}

\author{Piotr Czarnik}
\affiliation{\mbox{M. Smoluchowski Institute of Physics, Jagiellonian University,
             prof. S. {\L}ojasiewicza 11, PL-30348 Krak\'ow, Poland}}
\affiliation{\mbox{Institute of Nuclear Physics, Polish Academy of Sciences,
             Radzikowskiego 152, PL-31342 Krak\'ow, Poland}}

\author{Jacek Dziarmaga}
\affiliation{\mbox{M. Smoluchowski Institute of Physics, Jagiellonian University,
             prof. S. {\L}ojasiewicza 11, PL-30348 Krak\'ow, Poland}}

\author {     Andrzej M. Ole\'{s} }
\email[Author's email: ]{a.m.oles@fkf.mpg.de}
\affiliation{\mbox{M. Smoluchowski Institute of Physics, Jagiellonian University,
             prof. S. {\L}ojasiewicza 11, PL-30348 Krak\'ow, Poland}}
\affiliation{ Max Planck Institute for Solid State Research,
              Heisenbergstrasse 1, D-70569 Stuttgart, Germany }

\date{\today}

\begin{abstract}
We investigate thermodynamic phase transitions in the compass model and
in $e_g$ orbital model on an infinite square lattice by variational
tensor network renormalization (VTNR) in imaginary time.
The onset of nematic order in the quantum compass model is estimated
at ${\cal T}_c/J=0.0606(4)$.
For~the $e_g$ orbital model one finds:
($i$) a very accurate estimate of ${\cal T}_c/J=0.3566\pm 0.0001$ and
($ii$)~the~critical exponents in the Ising universality class.
Remarkably large difference in frustration results in so distinct
values of ${\cal T}_c$, while entanglement influences the quality
of ${\cal T}_c$ estimation.
\end{abstract}


\maketitle

\subsection{1. Introduction}
\label{sec:intro}

In a two-dimensional (2D) spin systems with exchange interactions
having SU(2) symmetry long-range order is excluded as stated by Mermin
and Wagner \cite{MeWa}. This paradigm fails in generic orbital models,
such as the 2D compass \cite{vdB15} or 2D $e_g$ orbital \cite{vdB99}
model, where interactions do not satisfy the assumptions
of the Mermin-Wagner theorem and the involved pseudospins order below
a thermodynamic phase transition at finite temperature ${\cal T}_c$.
Exchange interactions in both $e_g$ orbital and compass model may be
derived from the Ising model when the anisotropy and frustration of
exchange interactions between the two $a$ and $b$ axes in the square
lattice increases~\cite{Cin10}. Maximal frustration occurs in the 2D 
compass model and the nematic order is predicted by Quantum Monte 
Carlo (QMC) calculations below a rather low temperature 
${\cal T}_c=0.0585J$ \cite{Wen08}. The $e_g$ orbital model orders 
as well \cite{Ryn10} but the value of ${\cal T}_c$ is unknown as 
QMC calculations fail due to the sign problem.

Orbital models arise in a natural way when intraorbital Coulomb
interaction $U$ for partly filled $3d$ orbitals is large compared with
the hopping element $t$, electrons localize and the effective
interactions for a strongly correlated transition metal oxide with
orbital degeneracy are given by spin-orbital superexchange. Kugel and
Khomskii \cite{Kug82} were the first to recognize that spins and
orbitals are quantum and have to be treated on equal footing and their
interplay may lead to spectacular symmetry broken phases. Since then
various compounds are treated in detail and the field developed to
spin-orbital physics \cite{Tok00,Ole05,Kha05,Nor08,Ole12,Brz15}.
Spin-orbital models relevant for real materials are quite involved and
depend on the type of orbital degree of freedom, $e_g$ or $t_{2g}$.
While $e_g$ orbital interactions control the orbital order in
ferromagnetic planes of KCuF$_3$ and LaMnO$_3$ which is quite robust
and survives for spin disorder \cite{Rei05,Sna16}, compass interactions
stand for the pseudospin exchange in systems with strong spin-orbit
coupling at $4d$ (or $5d$) ions on a square lattice \cite{Jac09}.

The purpose of this paper is to present the results obtained with
Variational Tensor Network Renormalization (VTNR) in imaginary time
\cite{Cza15}. This algorithm approximates thermal density matrix
$\rho({\cal T})\propto e^{-H/\cal{T} }$ of a 2D quantum lattice model
with a Hamiltonian $H$ by a pair entangled projected operator (PEPO)
\cite{Wol08} which is an example of a tensor network \cite{Sch11}.
The refinement parameter is bond dimension $D$. Small $D$ is
sufficient to have exact $\rho({\cal T})$ representation for
a classical $H$ (e.g. $D=2$ for spins or pseudospins 1/2).
Larger $D$ enables VTNR to capture effects of quantum
fluctuations and entanglement \cite{Wol08}. In the limit $D\to \infty$
one can always recover non-truncated  $\rho({\cal T})$ although
polynomial scaling of CPU time with $D$ restricts numerical calculations
to $D\le 20$ in practice. The method like all tensor network methods
doesn't suffer from the sign problem which makes it powerful tool for
simulation of weakly entangled models.
VTNR was successfully applied to the 2D Hubbard model at ${\cal T}>0$
\cite{CRD16} and to estimate the value of ${\cal T}_c$ for both the
2D compass \cite{Cza16} and the $e_g$ orbital model \cite{Cza17}.

We discuss the nematic order in the 2D compass model in Sect. 2. Next
we present in Sect. 3 a very accurate estimate of ${\cal T}_c$ and the
critical exponents $\beta$ and $\gamma$ for the 2D $e_g$ orbital model.
The paper is summarized in Sect. 4 with a comparison between frustrated
Ising and two quantum models, $e_g$ orbital and compass.

\subsection{2. Nematic order in the compass model}
\label{sec:com}

The quantum compass model on a square lattice is \cite{vdB15},
\begin{equation}
H_{\rm com} = -\frac14 J \sum_j X_j X_{j+e_a}
              -\frac14 J \sum_j Z_j Z_{j+e_b}.
\label{H}
\end{equation}
Here $j$ is a site index, $X_j\equiv\sigma^x_j$ and
$Z_j\equiv\sigma^z_j$ are Pauli matrices at site $j$, and $e_a(e_b)$
are unit vectors along the $a(b)$ axis. As in the Ising model the sign
of $J$ in equation (\ref{H}) may be arbitrary and we take ferromagnetic
exchange $J=1$: $-\frac14 X_jX_{j+e_a}$ ($-\frac14 Z_jZ_{j+e_a}$)
for a bond along the $a(b)$ axis. The order parameter $Q>0$ is,
\beq
Q\equiv
\left|\left\langle Q_j\right\rangle_{\cal T}\right|=
\left|\left\langle X_j X_{j+e_a}-Z_j Z_{j+e_b}\right\rangle_{\cal T}\right|.
\label{OP}
\eeq
This order parameter is finite below the phase transition at 
${\cal T}_c$ --- the transition belongs to the 2D Ising universality 
class \cite{Wen10}. As expected for any ${\cal T}>0$ we find that local 
order parameters vanish within the numerical precision of $10^{-5}$, 
i.e., $\langle X_j\rangle_{\cal T}=\langle Z_j\rangle_{\cal T}=0$. 
Strong frustration implies that there is neither any local magnetization
nor any long-range order.

%%%%%%%%%%%%%%%%%%%%%%%%%%%%%%%%%%%%%%%%%%%%%%%%%%%%%%%%%%%%%%%%%%%%%%%%%%%%%%
\begin{figure}[t!]
\centering
\centering
\includegraphics[width=\columnwidth]{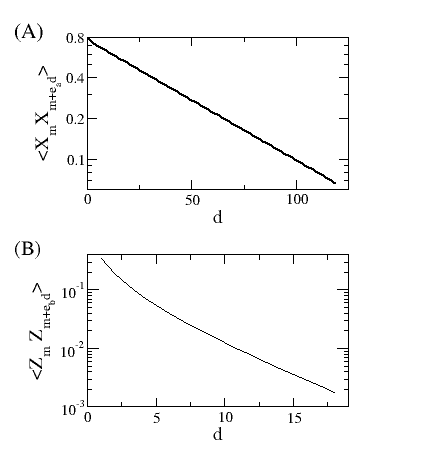}
\caption{Nematic order in the compass model at ${\cal T}=0.05814$:
Spin correlations for increasing distance $d$ along the $a(b)$ axis:
(A) the dominant correlation $\langle X_mX_{m+e_ad}\rangle$ \cite{Cza16},
and
(B)~the transverse correlation $\langle Z_mZ_{m+e_bd}\rangle$,
both for $D=15$.}
\label{fig:com}
\end{figure}
%%%%%%%%%%%%%%%%%%%%%%%%%%%%%%%%%%%%%%%%%%%%%%%%%%%%%%%%%%%%%%%%%%%%%%%%%%%%%%%%

For the phase transition to nematic order in the 2D compass model a
value of ${\cal T}_c=0.0585$ was estimated by QMC \cite{Wen10}, while
a high-temperature series expansion predicts a similar value
${\cal T}_c=0.0625$ \cite{Oit11}. The value of ${\cal T}_c$ is strongly
suppressed by quantum fluctuations from the classical compass model
\cite{Wen08}. The high-temperature extrapolation gives the
susceptibility exponent $\gamma\simeq 1.3$. Our estimate
$\gamma=1.35$ is in the same range and the order parameter
$Q({\cal T})$ exponent is $\beta=0.223$ \cite{Cza16}. Unfortunately the
convergence in $D$ is rather slow \cite{Cza16}, so these values are
close but not sufficiently close to the exact 2D Ising values,
$\gamma=\frac74$ and $\beta=\frac18$.

The nature of nematic order in the 2D quantum compass model is better
understood by studying its symmetries. The spectral properties can be
uniquely determined by discrete symmetries like parity. The conservation
of spin parities in rows and columns in the 2D quantum compass model
(for $X$ and $Z$ operators) uncovers very interesting hidden order with
two-dimer correlations \cite{Brz13}. The two-site correlations
(\ref{OP}) in the nematic phase are shown in Fig. \ref{fig:com}. The
dominant correlation function (here along the $a$ axis) is exponential
but relatively long-ranged with a correlation length estimated at
$\xi_{\parallel}=40(2)$ \cite{Cza16}. The transverse correlations are
weaker and decay nearly exponentially with a much shorter length
$\xi_{\perp}=6.9(4)$.

%%%%%%%%%%%%%%%%%%%%%%%%%%%%%%%%%%%%%%%%%%%%%%%%%%%%%%%%%%%%%%%%%%%%%%%%%%%%
\begin{figure}[t!]
\vspace{-0cm}
\includegraphics[width=\columnwidth,clip=true]{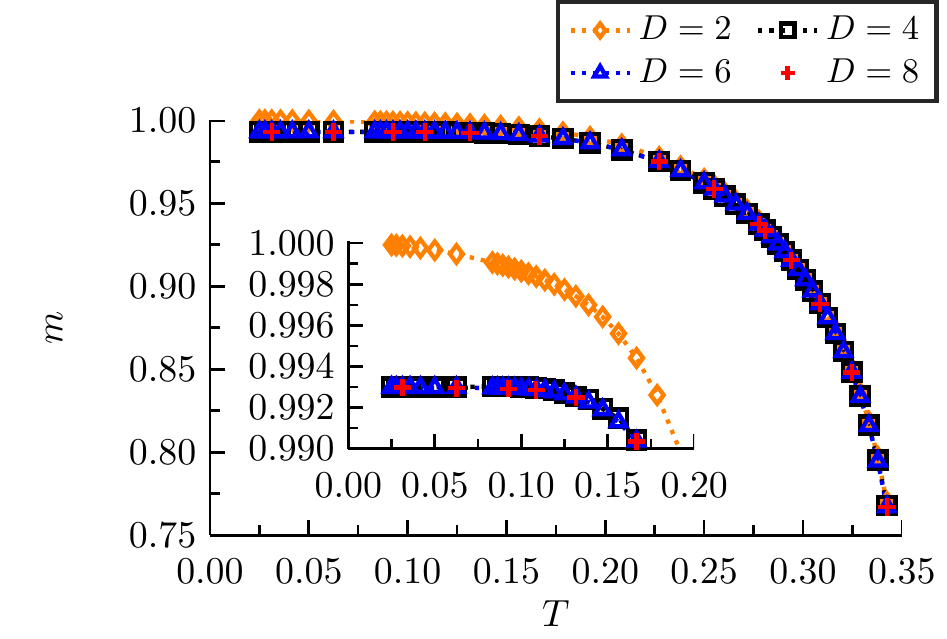}
\vspace{-0.3cm}
\caption{The order parameter $m({\cal T})$ (\ref{m}) as a function of
temperature ${\cal T}$ below the critical regime. The inset shows the
zoom of $m({\cal T})$ at low temperature ${\cal T}<0.18$.
The results demonstrate fast convergence in $D$ --- only $D=2$ gives
a higher $m({\cal T})$, while the data for $D=4,6,8$ all overlap.
}
\label{fig:eg}
\end{figure}
%%%%%%%%%%%%%%%%%%%%%%%%%%%%%%%%%%%%%%%%%%%%%%%%%%%%%%%%%%%%%%%%%%%%%%%%%%%%

\subsection{3. Ising-like order in the $e_g$ model}
\label{sec:com}

The quantum $e_g$ model on an infinite square lattice is obtained by
a proper transformation from the 2D Ising model \cite{Cin10} and is
defined by the Hamiltonian,
\begin{equation}
H_{eg} = - J\sum_j \tau^a_j \tau^a_{j+e_a}
       - J\sum_j \tau^b_j \tau^b_{j+e_b}.
\label{eg}
\end{equation}
The notation is analogous as in (1) with the coupling of different
orbital operators along the $a$ and $b$ axis:
\bea
\tau^a_j = \frac14\left( -\sigma^z_j + \sqrt3 \sigma^x_j \right),~~
\tau^b_j = \frac14\left( -\sigma^z_j - \sqrt3 \sigma^x_j \right).
\eea
As in Sect. 2, we take the isotropic model with $J=1$.

At low temperature a spontaneous breaking of symmetry is determined by
the term $\propto\frac{3}{16}\sigma^x_i\sigma^x_j$ \cite{Cin10}.
This symmetry breaking implies a finite order parameter
\beq
m({\cal T})\equiv\langle\sigma^x_j\rangle_{\cal T}.
\label{m}
\eeq
It could be obtained by a systematic convergence test in the bond
dimension $D$. Indeed, from $D=4$ the data practically fall on each
other and $D=8$ corresponds to the converged results, see Fig.
\ref{fig:eg}. The entanglement at ${\cal T}\to 0$ is small and
therefore $m({\cal T})$ converges in $D$ so fast. The ground state
order parameter is almost saturated,
$m(0)=0.993$,
in agreement with the Multiscale Entanglement Renormalization Ansatz
(MERA) \cite{Cin10}, and the quantum fluctuations are very weak at
${\cal T}=0$ \cite{vdB99}.

Unlike the 2D compass model \cite{Wen10}, the $e_g$ orbital model
(\ref{eg}) is not tractable by QMC \cite{Wen11}, but the order
parameter $m({\cal T})$ and the susceptibility $\chi({\cal T})$ were
found by the VTNR \textit{Anzatz}. Their behavior near ${\cal T}_c$ is:
\beq
\label{mT}
   m({\cal T})\propto({\cal T}_c-{\cal T})^{\beta}, \hskip .7cm
\chi({\cal T})\propto({\cal T}-{\cal T}_c)^{\gamma}.
\eeq
The convergence in $D$ is fast \cite{Cza17} and one finds:
$\beta\simeq 0.126$ and $\gamma\simeq 1.736$ for $D\ge 7$;
they approach the critical exponents of the 2D Ising model,
$\beta=\frac18$ and $\gamma=\frac74$.
From the convergence of $m({\cal T})$ and $\chi({\cal T})$ (\ref{mT})
we deduce ${\cal T}_c\simeq 0.3566$ \cite{Cza17}. Note that from the
leading $\propto\frac{3}{16}\sigma^x_i\sigma^x_j$ term in (\ref{eg})
and the Ising model \cite{Lars} one expects
${\cal T}_c\simeq 0.4255$ but the quantum fluctuations are activated by
temperature and reduce the above value of ${\cal T}_c$ by $\sim 16$\%.

%%*******************************************
\begin{table}[t!]
\centering
\caption{The critical temperature ${\cal T}_c$ and the type of order
for the classical and quantum models on a square lattice:
Ising model, $\frac12$ and $\frac23$ frustrated Ising \cite{Lon80},
fully frustrated Villain model \cite{Vil77}, $e_g$ orbital model
\cite{Cza17} and 2D compass model~\cite{Cza16}.
}
\begin{ruledtabular}
\begin{tabular}{ccccc}
 2D model            &  order &   ${\cal T}_c/J$   &  method &   citation   \cr \hline
    Ising            &   2D   &      0.567296      &  exact  & \cite{Lars}  \cr
$\frac12$ frustrated &   2D   &      0.410         &  exact  & \cite{Lon80} \cr
$\frac23$ frustrated &   2D   &      0.342         &  exact  & \cite{Lon80} \cr
 Villain             &  ---   &       0.0          &  exact  & \cite{Vil77} \cr \hline
 $e_g$ orbital       &   2D   & $0.3566\pm 0.0001$ &  VTNR   & \cite{Cza17} \cr
 compass            & nematic & $0.0606\pm 0.0004$ &  VTNR   & \cite{Cza16} \cr
\end{tabular}
\end{ruledtabular}
\label{tab:Tc}
\end{table}
%%*******************************************

\subsection{4. Summary}
\label{sec:com}

To highlight the difference between geometrical frustration
\cite{Mat06} in classical and intrinsic frustration \cite{Ole12} in
quantum models, we compare the values of ${\cal T}_c$ for the 2D $e_g$
orbital and compass models with those for 2D frustrated Ising model in
Table I. Increasing the number of frustrated plaquettes on a square
lattice reduces ${\cal T}_c$ but as long as ladders of non-frustrated
plaquettes exist, the decrease of ${\cal T}_c$ is slow \cite{Lon80},
and only when all the plaquettes are frustrated the 2D order totally
collapses in the Villain model \cite{Vil77}.
In contrast, for complete quantum frustration in the compass model,
the ground state is highly degenerate but a novel nematic order
emerges below ${\cal T}_c>0$. For partial frustration in the $e_g$
orbital model the reduction of ${\cal T}_c$ is more dramatic as
quantum fluctuations are activated in the critical regime.

To summarize, by comparing the $e_g$ orbital and quantum compass model
we conclude that the VTNR \textit{Ansatz} is particularly efficient
when entanglement is weak. Then a very accurate treatment is possible,
even when the system suffers from the fermionic sign problem.
We suggest that both the 2D $e_g$ and compass model are in the
2D Ising universality class and present a very accurate estimate of
${\cal T}_c$ for the less entangled $e_g$ model.

\subsection{Acknowledgments}
\vskip -.3cm

We kindly acknowledge support by Narodowe Centrum Nauki
(NCN) under Projects
No.~2016/23/B/ST3/00830 (P.C. and J.D.),
No.~2016/23/B/ST3/00839 (A.M.O.), and
No.~2015/16/T/ST3/00502 (Ph.D. thesis of P.C.).

%[1]   DOI:\\ http://dx.doi.org/10.1070/PU1982v025n04ABEH004537


\vskip -.3cm
\begin{thebibliography}{99}

%intro
\bibitem{MeWa}  N.D. Mermin, H. Wagner,
                   \textit{Phys. Rev. Lett.\/} \textbf{17}, 1133 (1966).
\\ DOI: 10.1103/PhysRevLett.17.1133

\bibitem{vdB15} Z. Nussinov, J. van den Brink,
                   \textit{Rev. Mod. Phys.\/} \textbf{86}, 1 (2015).
\\ DOI: 10.1103/RevModPhys.87.1

\bibitem{vdB99} J. van den Brink, P. Horsch, F. Mack, A.M. Ole\'s,
                   \textit{Phys. Rev. B} \textbf{59}, 6795 (1999).
\\ DOI: 10.1103/PhysRevB.59.6795

%compass
\bibitem{Cin10} L. Cincio, J. Dziarmaga, A.M. Ole\'s,
                   \textit{Phys. Rev. B\/} \textbf{82}, 104416 (2010).
\\ DOI: 10.1103/PhysRevB.82.104416

\bibitem{Wen08} S. Wenzel, W. Janke,
                   \textit{Phys. Rev. B\/} \textbf{78}, 064402 (2008).
\\ DOI: 10.1103/PhysRevB.78.064402

\bibitem{Ryn10} A. van Rynbach, S. Todo, S. Trebst,
                   \textit{Phys. Rev. Lett.\/} \textbf{105}, 146402 (2010).
\\ DOI: 10.1103/PhysRevLett.105.146402

\bibitem{Kug82} K.I. Kugel, D.I. Khomskii,
                   \textit{Sov. Phys. Usp.} \textbf{25}, 231 (1982).
\\ DOI: 10.1070/PU1982v025n04ABEH004537

\bibitem{Tok00} Y. Tokura, N. Nagaosa,
                   \textit{Science} \textbf{288}, 462 (2000).
\\ DOI: 10.1126/science.288.5465.462

\bibitem{Ole05} A.M. Ole\'s, G. Khaliullin, P. Horsch, L.F. Feiner,
                   \textit{Phys. Rev. B\/} \textbf{72}, 214431 (2005).
\\ DOI: 10.1103/PhysRevB.72.214431

\bibitem{Kha05} G. Khaliullin,
                   \textit{Prog. Theor. Phys. Suppl.\/} \textbf{160}, 155 (2005).
\\ DOI: 10.1143/PTPS.160.155

%som -- entanglement
\bibitem{Nor08} B. Normand, A.M. Ole\'s,
                   \textit{Phys. Rev. B\/} \textbf{78}, 094427 (2008);\\
                B. Normand,
                   \textit{Phys. Rev. B\/} \textbf{83}, 064413 (2011);
                J. Chaloupka, A.M. Ole\'s,
                   \textit{Phys. Rev. B\/} \textbf{83}, 094406 (2011).
\\ DOI: 10.1103/PhysRevB.78.094427
\\ DOI: 10.1103/PhysRevB.83.064413
\\ DOI: 10.1103/PhysRevB.83.094406

\bibitem{Ole12} A.M. Ole\'s,
                   \textit{J. Phys.: Condens. Mat.\/} \textbf{24}, 313201 (2012);
                   \textit{Acta Phys. Pol. A} \textbf{127}, 163 (2015).
\\ DOI: 10.1088/0953-8984/24/31/313201
\\ http://przyrbwn.icm.edu.pl/APP/ABSTR/127/a127-2-4.html
\\ ( please insert DOI )

\bibitem{Brz15} W. Brzezicki, A.M. Ole\'s, M. Cuoco,
                   \textit{Phys. Rev. X} \textbf{5}, 011037 (2015);
                W. Brzezicki, M. Cuoco, A.M.~Ole\'s,
                   \textit{J.~Sup. Novel Magn.} \textbf{29}, 563 (2016);
                                                \textbf{30}, 129 (2017).
\\ DOI: 10.1103/PhysRevX.5.011037
\\ DOI: 10.1007/s10948-015-3287-z
\\ DOI: 10.1007/s10948-016-3750-5

\bibitem{Rei05} A.J.W. Reitsma, L.F. Feiner, A.M. Ole\'s,
                   \textit{New J. Phys.\/} \textbf{7}, 121 (2005).
\\ DOI: 10.1088/1367-2630/7/1/121

\bibitem{Sna16} M. Snamina, A.M. Ole\'s,
                   \textit{Phys. Rev. B\/} \textbf{94}, 214426 (2016).
\\ DOI: 10.1103/PhysRevB.94.214426

\bibitem{Jac09} G. Jackeli, G. Khaliullin,
                   \textit{Phys. Rev. Lett.\/} \textbf{102}, 017205 (2009).
\\ DOI: 10.1103/PhysRevLett.102.017205

%som -- entanglement
\bibitem{Cza15} P. Czarnik, J. Dziarmaga,
                   \textit{Phys. Rev. B} \textbf{92}, 035120 (2015);
                                         \textbf{92}, 035152 (2015).
\\ DOI: 10.1103/PhysRevB.92.035120
\\ DOI: 10.1103/PhysRevB.92.035152

\bibitem{Wol08} M.M. Wolf, F. Verstraete, M.B. Hastings, J.I. Cirac,
                   \textit{Phys. Rev. Lett.} \textbf{100}, 070502.
\\ DOI: 10.1103/PhysRevLett.100.070502

\bibitem{Sch11} U. Schollw\"ock,
                   \textit{Ann. of Phys.}, \textbf{326}, 96 (2011).
\\ DOI: 10.1016/j.aop.2010.09.012

\bibitem{CRD16} P. Czarnik, M.M. Rams, J. Dziarmaga,
                   \textit{Phys. Rev. B} \textbf{94}, 235142 (2016).
\\ DOI: 10.1103/PhysRevB.94.235142

%eg
\bibitem{Cza16} P. Czarnik, J. Dziarmaga, A.M. Ole\'s,
                   \textit{Phys. Rev. B} \textbf{93}, 184410 (2016).
\\ DOI: 10.1103/PhysRevB.93.184410

\bibitem{Cza17} P. Czarnik, J. Dziarmaga, A.M. Ole\'s,
                   \textit{Phys. Rev. B} \textbf{96}, 014410 (2017).
\\ DOI: 10.1103/PhysRevB.96.014410

\bibitem{Wen10} S. Wenzel, W. Janke, A.M. L\"auchli,
                   \textit{Phys. Rev. E} \textbf{81}, 066702 (2010).
\\ DOI: 10.1103/PhysRevE.81.066702

\bibitem{Oit11} J. Oitmaa, C.J. Hamer,
                   \textit{Phys. Rev. B} \textbf{83}, 094437 (2011).
\\ DOI: 10.1103/PhysRevB.87.214421

\bibitem{Brz13} W. Brzezicki, A.M. Ole\'s,
                   \textit{Phys. Rev. B\/} \textbf{82}, 060401 (2010);
                                           \textbf{87}, 214421 (2013).
\\ DOI: 10.1103/PhysRevB.82.060401
\\ DOI: 10.1103/PhysRevB.87.214421

\bibitem{Wen11} S. Wenzel, A.M. L\"auchli,
                   \textit{Phys. Rev. Lett.} \textbf{106}, 197201 (2011).
\\ DOI: 10.1103/PhysRevLett.106.197201

\bibitem{Lars}  Lars Onsager,
                   \textit{Phys. Rev.} \textbf{65}, 117 (1944).
\\ DOI: 10.1103/PhysRev.65.117

\bibitem{Mat06} D.C. Mattis,
                   \textit{The Theory of Magnetism Made Simple},
                   World Scientific, New Jersey, 2006.

\bibitem{Lon80} L. Longa, A.M. Ole\'s,
                   \textit{J. Phys. A: Math. Theor.} \textbf{13}, 1031 (1980).
\\ DOI: 10.1088/0305-4470/13/3/036

\bibitem{Vil77} J. Villain,
                   \textit{J. Phys. C: Solid State Phys.} \textbf{10}, 1717 (1977).
\\ DOI: 10.1088/0022-3719/10/10/014



\end{thebibliography}
\end{document}